# The quantum condition space

Zixuan Hu and Sabre Kais*

*Department of Chemistry, Department of Physics, and Purdue Quantum Science and Engineering Institute, Purdue University, West Lafayette, IN 47907, United States*
*Email:* kais@purdue.edu

In this work we first propose to exploit the fundamental properties of quantum physics to evaluate the probability of events with projection measurements. Next, to study what events can be specified by quantum methods, we introduce the concept of the *condition space*, which is found to be the dual space of the classical outcome space of bit strings. Just like the classical outcome space generates the quantum state space, the condition space generates the *quantum condition space* that is the central idea of this work. The quantum condition space permits the existence of entangled conditions that have no classical equivalent. In addition, the quantum condition space is related to the quantum state space by a Fourier transform guaranteed by the Pontryagin duality, and therefore an entropic uncertainty principle can be defined. The quantum condition space offers a novel perspective of understanding quantum states with the duality picture. In addition, the quantum conditions have physical meanings and realizations of their own and thus may be studied for purposes beyond the original motivation of characterizing events for probability evaluation. Finally, the relation between the condition space and quantum circuits provides insights into how quantum states are collectively modified by quantum gates, which may lead to deeper understanding of the complexity of quantum circuits.

## 1. Introduction

Quantum computation has achieved enormous progress in the last 20 years [1-7] with state-of-the-art technologies emerging one after another [8-13]. Over the years, numerous quantum algorithms have been proposed with the potential to outperform classical methods with decisive quantum advantage – these include the phase estimation algorithm [14], Shor's factorization algorithm [15], the Harrow-Hassidim-Lloyd algorithm for linear systems [16], the hybrid classical-quantum algorithms [17, 18], the quantum machine learning algorithms [19, 20], and quantum algorithms for open quantum dynamics [21-23].

In this work we first propose the idea of utilizing fundamental properties of quantum physics to evaluate the probability of events with projection measurements. Representing a large probability distribution with a qubit-based quantum state vector and evaluating event probability by projecting into a subspace, a decisive advantage over classical methods can be achieved. We then proceed to study what kinds of events can be specified by the so-called half-set conditions such that they can be easily realized by quantum gates. This leads us to the *condition space* that is found to be the dual space of the classical outcome space of bit strings. Next, just like the quantum state space can be generated by using members of the outcome space as basis states, *the quantum condition space* can also be generated by using members of the condition space as basis conditions. The quantum condition space is the central idea of this work, and we design a formalism for it based on a creation called the *"q-condition"* that serves a similar role to the qubit in the usual quantum state space. With the help of the "q-condition", the quantum condition vector can be physically interpreted and a potential realization with the usual tools of quantum circuits is proposed. Notably, the quantum condition space permits the existence of *entangled conditions* that have no classical equivalent. The quantum condition space and the quantum state space are related through the mathematical concept of Pontryagin duality [24], which permits a Fourier transform and an entropic uncertainty principle between the two – this is analogous to the relation between the position space and the momentum space in fundamental quantum physics. The quantum condition space deepens our understanding of the quantum state space with the duality picture. In addition, the quantum conditions have physical meanings and realizations, and thus may deserve further attention as physical objects independent from the quantum states. Finally, going beyond the original motivation of evaluating event probability, we discuss a deeper relation between the half-set conditions and quantum circuits, which leads to future directions in better understanding the complexity of quantum circuits.

## 2. Evaluating event probability with quantum gates and measurements

To motivate the study on the quantum condition space, we start with the task of evaluating event probability with quantum gates and measurements. Consider an *n*-qubit quantum state vector with $2^n$ complex number entries $C_i$. With the normalization condition $\sum_{i=0}^{2^n-1}|C_i|^2 = 1$, the quantum state vector naturally corresponds to a discrete and finite probability distribution with the probability $|C_i|^2$ associated with the $i^{th}$ basis state: e.g. for a 3-qubit case, $|C_0|^2 = |C_{000}|^2$ corresponds to $|000\rangle$, $|C_3|^2 = |C_{011}|^2$ corresponds to $|011\rangle$, etc. Here in terms of probability theory, each basis state can be considered as an outcome. To evaluate the probability associated with any particular outcome we can project into the corresponding basis state $|i\rangle$ (*i* a binary integer) to statistically measure $|C_i|^2$. In probability theory it is often required to evaluate the probability of events that are collections of large number of outcomes. Classically given a discrete and finite probability distribution, the probability of an event is calculated by summing over the probabilities of all the outcomes included in the event. When the number of outcomes scales exponentially with the qubit number *n*, this task becomes computationally expensive with large *n*. Now exploiting the fundamental property of quantum measurement, we can efficiently evaluate the probability of certain events containing exponentially large number of outcomes. For example, projecting the *n*-qubit quantum state vector into the first qubit subspace with the condition "$q_1 = 0$", we can evaluate the sum $\sum_{q_1=0}|C_i|^2$ by measuring the statistics of the outcomes satisfying "$q_1 = 0$" versus those satisfying "$q_1 = 1$". This is essentially a sampling procedure for which multiple projection measurements are required. A basic result in statistics says that if we use a finite number of sampling measurements to reproduce the exact probability distribution, then the error of the sampling decreases with increasing number of measurements. In particular, if the error of the sampling can be represented by the standard error of the mean $\sigma_{mean}$, and the exact probability distribution defines a standard deviation $\sigma$, then:

$$\sigma_{mean} = \frac{\sigma}{\sqrt{P}}, \qquad P = \left(\frac{\sigma}{\sigma_{mean}}\right)^2 \tag{1}$$

where $P$ is the number of measurements. Here we see that $P$ does not scale with the number of qubits $n$, but only depends on the error $\sigma_{mean}$ we choose to tolerate. Now back to the example with the event specified by the condition "$q_1 = 0$", there are $2^{n-1}$ outcomes – half of the total of $2^n$ outcomes – in the event, and thus it is exponentially expensive to sum over the outcome probabilities by classical methods. In the meanwhile, the number $P$ of quantum projection measurements needed is a constant in the qubit number $n$, thus as long as the initiation of the quantum state vector has a polynomial complexity of $O(poly(n))$, then the total complexity of quantum evaluation of the event probability will be polynomial. This is indeed possible for any quantum states that have polynomial complexity, such as those that have been characterized as the "standard states" in our previous work on quantum state complexity [25]:

$$\begin{aligned}&\left[\left(a_1|00\rangle_{12} + a_2|11\rangle_{12}\right)b_1|0\rangle_3 + \left(a_1|01\rangle_{12} + a_2|10\rangle_{12}\right)b_2|1\rangle_3\right]c_1|0\rangle_4 \\ &+ \left[\left(a_1|00\rangle_{12} + a_2|11\rangle_{12}\right)b_1|1\rangle_3 + \left(a_1|01\rangle_{12} + a_2|10\rangle_{12}\right)b_2|0\rangle_3\right]c_2|1\rangle_4\end{aligned} \tag{2}$$

Shown in Equation (2) is a 4-qubit "minimal standard state" whose quantum state vector can be efficiently initiated by the procedure described in Ref. [25]. The coefficients $a_i$, $b_i$, $c_i$ are arbitrary complex numbers that satisfy $|a_1|^2 + |a_2|^2 = 1$, $|b_1|^2 + |b_2|^2 = 1$, and $|c_1|^2 + |c_2|^2 = 1$. Because this quantum state has all the qubits entangled, there can be no classical equivalent to represent the corresponding probability distribution as efficiently as the quantum state does. Consequently, when the number of qubits $n$ becomes large, the aforementioned quantum evaluation of the probability of the event specified by the condition "$q_1 = 0$" will have decisive polynomial-versus-exponential advantage over any classical methods.

Now a natural question to ask is in addition to the condition "$q_1 = 0$", what are other conditions that can be used to specify an event such that its probability can be efficiently evaluated? Obviously the condition associated with each qubit "$q_i = 0$" and its complement "$q_i = 1$" each specifies a different half-set of $2^{n-1}$ outcomes. In addition, the conditions associated with arbitrary sums of the qubits, e.g. "$q_1 \oplus q_2 = 0$" ($\oplus$ here means addition modulo 2), "$q_2 \oplus q_3 \oplus q_5 = 0$", etc., and their complements each specifies a different half-set of $2^{n-1}$ outcomes. The collection of all these half-set conditions are the building blocks of a Boolean algebra that includes all the conditions formed by applying "and", "or", and "not" operations on the half-set conditions. Note that any single outcome (e.g. 101) can be specified by conjunctions ("and") over half-set conditions defined with a single qubit (e.g. 101 is specified by "$q_1 = 1$ and $q_2 = 0$ and $q_3 = 1$"), so the Boolean algebra includes all possible events out of the $2^n$ outcomes, because any possible subset can be formed by unions of single outcomes and thus can be specified by disjunctions ("or") over the conditions for single outcomes. Of course, using disjunctions over single outcome conditions to specify an event is highly inefficient, thus in practice we want to directly work with the half-set conditions to build more complex conditions for specifying more complex events. Obviously the single qubit conditions "$q_i = 0$" and their complements "$q_i = 1$" can be directly evaluated by projection measurements. The other half-set conditions formed by sums of the qubits, e.g. "$q_1 \oplus q_2 = 0$" and "$q_2 \oplus q_3 \oplus q_5 = 0$", can also be efficiently evaluated by quantum gates and measurements. For "$q_1 \oplus q_2 = 0$", given a quantum state vector represented by qubits, apply $\text{CNOT}_{1 \to 2}$ (where $1 \to 2$ means $q_1$ controls $q_2$) and the value of $q_1 \oplus q_2$ will be stored on $q_2$ (if $\text{CNOT}_{2 \to 1}$ then the value will be stored on $q_1$), which can be seen from the truth table of the CNOT gate. Now measuring $q_2$ will allow us to realize the condition "$q_1 \oplus q_2 = 0$" or its complement "$q_1 \oplus q_2 = 1$". Similarly the example "$q_2 \oplus q_3 \oplus q_5 = 0$" can be realized by applying $\text{CNOT}_{2 \to 3}$, then $\text{CNOT}_{3 \to 5}$, and finally measuring $q_5$. In general, any other conditions formed by sums of arbitrary qubits can be realized by applying at most $n-1$ CNOT gates and then measuring the target qubit of the final CNOT. This result justifies our choice of the half-set conditions as the building blocks for more complex conditions, because any half-set conditions can be easily realized by $O(n)$ CNOT gates, which are themselves the basic operations of quantum computing. Next we define the *polynomial conditions* to be those that can be formed by a polynomial number (a number polynomially scaled with *n*) of Boolean operations on half-set conditions. The events specified by polynomial conditions are *polynomial events* that can be efficiently evaluated by quantum gates and measurements, given that the quantum state vector representing the probability distribution can be efficiently initialized: e.g. the state vector in Equation (2). As previous examples of the half-set conditions show, many of these polynomial events contain exponential number of outcomes and are hard to evaluate by classical means, given that the vector representing the probability distribution cannot be efficiently represented by classical systems: e.g. the state vector in Equation (2). To identify and characterize the polynomial events then becomes important because these events may allow us to demonstrate decisive quantum advantage for the task of event probability evaluation.

### 3. Theory of the quantum condition space

#### 3.1 The condition space as the dual of the outcome space.

So far we have seen the motivation for studying the half-set conditions that allow efficient quantum evaluation of the probabilities of events. To deeper understand the half-set conditions, we first take the collection of all the "0" conditions – those half-set conditions with "0" on the right side – and notice a correspondence between the outcomes and the conditions:

Without loss of generality, take the 3-qubit outcome 001 as an example, it satisfies the following "0" conditions: "$q_1 = 0$", "$q_2 = 0$", "$q_1 \oplus q_2 = 0$". In the meanwhile the condition "$q_3 = 0$" is satisfied by the following outcomes: 100, 010, 110, 000. Note the similarity in the forms of 001 and "$q_3 = 0$", 100 and "$q_1 = 0$", 010 and "$q_2 = 0$", 110 and "$q_1 \oplus q_2 = 0$": i.e. whenever "1" appears for $q_i$ in an outcome, we can find a corresponding "0" condition with $q_i$ contributing to the sum on the left side. Now if we define 000 to correspond to the condition "$0 = 0$" (the always-true condition), then there is a one-to-one correspondence between the conditions satisfied by 001 (now also including "$0 = 0$") and the outcomes satisfying "$q_3 = 0$".

This correspondence between the outcomes and the conditions also applies to all other outcomes and conditions, and we now show that it is not a coincidence. Again using the 3-qubit case without loss of generality, the collection of all outcomes can be understood as a 3-dimensional linear space $V$ over the binary field $\{0,1\}$ and any outcome can be expressed as a linear combination of the three basis vectors (100), (010), and (001): $\mathbf{v} = (q_1, q_2, q_3) = q_1(100) + q_2(010) + q_3(001)$, where the coefficients $q_i$'s take values 0 or 1, and the addition "+" is bit-wise addition modulo 2. From the theory of linear spaces, the collection of all the linear functionals over the space $V$: $f(\mathbf{v}) = f((q_1, q_2, q_3)) = f_1 q_1 \oplus f_2 q_2 \oplus f_3 q_3$, forms the linear space $V^*$ that is the *dual* of $V$. Here the three basis functionals are $f^{(1)}((q_1,q_2,q_3)) = q_1$, $f^{(2)}((q_1,q_2,q_3)) = q_2$, $f^{(3)}((q_1,q_2,q_3)) = q_3$, the coefficients $f_i$'s take values 0 or 1, and the "$\oplus$" is addition modulo 2. Clearly, $V$ and $V^*$ have the same dimensions, and in our finite case the same number of vectors. Now note that for any linear functional $f((q_1, q_2, q_3)) = f_1 q_1 \oplus f_2 q_2 \oplus f_3 q_3$, we can associate it with a half-set "0" condition "$f_1 q_1 \oplus f_2 q_2 \oplus f_3 q_3 = 0$" and then define a linear space for the conditions: the three basis conditions are "$q_1 = 0$", "$q_2 = 0$", "$q_3 = 0$" and any linear combination of these is defined as:

$$f_1 \cdot "q_1 = 0" \oplus f_2 \cdot "q_2 = 0" \oplus f_3 \cdot "q_3 = 0" \xrightarrow{\text{defined as}} "f_1 q_1 \oplus f_2 q_2 \oplus f_3 q_3 = 0" \quad (3)$$

where we see the addition and scalar multiplication on the conditions are defined such that the items on the left sides of the conditions calculate with the same rule for the functional space, and the result condition is obtained by attaching "$= 0$" to the result functional. Consequently the half-set "0" conditions form a linear space with exactly the same properties of the linear functional space $V^*$ and thus this *condition space* becomes the dual of the outcome space.

Now the previously mentioned one-to-one correspondence between the outcome space and the condition space can be easily explained by the relation between the linear space $V$ and its dual $V^*$. First a natural relation between $V$ and $V^*$ is given by the orthogonality condition $f_1 q_1 \oplus f_2 q_2 \oplus f_3 q_3 = 0$: in the example above, the outcome (001) means $q_1 = 0$, $q_2 = 0$, and $q_3 = 1$, thus by the orthogonality condition $f_3 = 0$, which means the conditions belonging to the subspace "$f_1 q_1 \oplus f_2 q_2 = 0$" are related to (001). Indeed this includes the conditions "$q_1 = 0$", "$q_2 = 0$",

"$q_1 \oplus q_2 = 0$", and "$0 = 0$", which are all the conditions satisfied by (001). Now by the theory of dual spaces, the duality between $V$ and $V^*$ is mutual (the dual of $V^*$ is $V$), so there must also be a condition in $V^*$ playing the equivalent role of (001) in $V$, that is "$q_3 = 0$", which means $f_1 = 0$, $f_2 = 0$, and $f_3 = 1$. So by the same orthogonality condition we find the outcomes satisfying "$q_3 = 0$" are (100), (010), (110), and (000), corresponding respectively to "$q_1 = 0$", "$q_2 = 0$", "$q_1 \oplus q_2 = 0$", and "$0 = 0$". By the same reasoning, we have the fact that any vector in the outcome space has a one-to-one correspondence to a unique vector in the condition space. All the results in this section – the duality between outcome space and the condition space, the correspondence between them, and the natural relation given by the orthogonality condition – can be easily generalized to an $n$-qubit case.

### 3.2 The quantum condition space

Now the duality between the outcome space and the condition space has been established, we proceed to introduce the central idea of this study: the *quantum condition space*. So far in our discussion, vectors in the 3-qubit outcome space such as (100) and (110) have been classical entities, but they can also be considered as basis states $|100\rangle$ and $|110\rangle$ of the quantum state space. Indeed, the quantum state space is the usual Hilbert space we are familiar with: e.g. the 3-qubit quantum state space has $2^3 = 8$ dimensions, the vectors in the 3-qubit outcome space $|000\rangle$, $|001\rangle$, $|010\rangle$ ..., $|111\rangle$ are basis states, and the scalars are complex numbers. However, now the basis states have been assigned the additional meaning as vectors in the outcome space, it is logical to ask the question: what would happen if the vectors of the dual space of the outcome space, i.e. vectors of the condition space, are used to build a Hilbert space? The answer is the *quantum condition space*. Formally we can define the vectors of the condition space as the basis vectors: e.g. $|000]$, $|001]$, $|010]$, ..., $|111]$, and then any member of the quantum condition space will be:

$$\phi = \sum_{i=0}^{2^n-1} D_i |i], \quad \text{with} \sum_{i=0}^{2^n-1} |D_i|^2 = 1, \quad i \text{ is a binary integer and } D_i \text{ is a complex number} \quad (4)$$

where we have used the symbol $| \cdot ]$ to distinguish the quantum condition basis from the quantum state basis. To interpret the formal definition physically, recall that for the usual quantum state vector $\psi = \sum_{i=0}^{2^n-1} C_i |i\rangle$, we interpret it by breaking it down into 1-qubit superposition states such as $|u\rangle = a_1 |0\rangle + a_2 |1\rangle$ or 2-qubit entangled states such as $b_1 |u_1 v_1\rangle + b_2 |u_2 v_2\rangle$, where $|a_1|^2 + |a_2|^2 = 1$, $|b_1|^2 + |b_2|^2 = 1$; $u_1$, $u_2$, $v_1$ and $v_2$ are 1-qubit states with $\langle u_1 | u_2 \rangle = 0$ and $\langle v_1 | v_2 \rangle = 0$ (for how an arbitrary quantum state can be broken into 1-qubit and 2-qubit states, see Ref. [25]). In other words, the $\psi = \sum_{i=0}^{2^n-1} C_i |i\rangle$ is a mathematical form, and it is the superposition $|u\rangle = a_1 |0\rangle + a_2 |1\rangle$ or the entanglement $b_1|u_1v_1\rangle+b_2|u_2v_2\rangle$ that gives the physical meaning in terms of how the qubits behave in this quantum state. Here in a similar way, we define superposition in the quantum condition space as:

$$|u] = a_1|0] + a_2|1] \tag{5}$$

and the *product condition* is (analogous to the product state):

$$\begin{aligned}|uv]_{12} &= |u]_1 \oplus |v]_2 \\ &= (a_1|0]_1 + a_2|1]_1) \oplus (b_1|0]_2 + b_2|1]_2) \\ &= a_1b_1(|0]_1 \oplus |0]_2) + a_1b_2(|0]_1 \oplus |1]_2) + a_2b_1(|1]_1 \oplus |0]_2) + a_2b_2(|1]_1 \oplus |1]_2) \\ &= a_1b_1|00] + a_1b_2|01] + a_2b_1|10] + a_2b_2|11]\end{aligned} \tag{6}$$

where $\oplus$ behaves like multiplication between $|u]_1$ and $|v]_2$ in the quantum condition space, while $+$ plays the role of vector addition. Note $\oplus$ still retains the meaning of addition modulo 2 in the condition space. The usual quantum product state is:

$$\begin{aligned}|uv\rangle_{12} &= |u\rangle_1 \otimes |v\rangle_2 \\ &= (a_1|0\rangle_1 + a_2|1\rangle_1) \otimes (b_1|0\rangle_2 + b_2|1\rangle_2) \\ &= a_1b_1(|0\rangle_1 \otimes |0\rangle_2) + a_1b_2(|0\rangle_1 \otimes |1\rangle_2) + a_2b_1(|1\rangle_1 \otimes |0\rangle_2) + a_2b_2(|1\rangle_1 \otimes |1\rangle_2) \\ &= a_1b_1|00\rangle + a_1b_2|01\rangle + a_2b_1|10\rangle + a_2b_2|11\rangle\end{aligned} \tag{7}$$

Comparing Equations (6) and (7) we see that by using $\oplus$ in place of $\otimes$ like a multiplication, a product condition in the quantum condition space have the same forms and properties as a product state in the quantum state space. Now we can physically interpret the product condition as $|00] = |0]_1 \oplus |0]_2$ means the always-true condition "$0 = 0$", $|01] = |0]_1 \oplus |1]_2$ means the condition "$q_2 = 0$", $|10] = |1]_1 \oplus |0]_2$ means the condition "$q_1 = 0$", and $|11] = |1]_1 \oplus |1]_2$ means the condition "$q_1 \oplus q_2 = 0$". In general, the expression $|0]_i$ means $q_i$ is *missing* from the left side of the condition, and $|1]_i$ means $q_i$ is *present* at the left side of the condition, e.g. $|101] = |1]_1 \oplus |0]_2 \oplus |1]_3$ means "$q_1 \oplus q_3 = 0$" and $|10110] = |1]_1 \oplus |0]_2 \oplus |1]_3 \oplus |1]_4 \oplus |0]_5$ means "$q_1 \oplus q_3 \oplus q_4 = 0$". The entity $|u]_i = a_1|0]_i + a_2|1]_i$ thus means $q_i$ is *in the superposition of being missing and present* – a typical quantum interpretation. $|u]_i$ plays the same foundational role of the 1-qubit state $|u\rangle_i$ and may be called the "*1-q-condition*". Similarly $|uv] = |u]_1 \oplus |v]_2$ is a 2-q-condition and $|uvw] = |u]_1 \oplus |v]_2 \oplus |w]_3$ is a 3-q-condition. Note that in certain situations it may be beneficial to explicitly indicate the missing qubits such that $|11] = |1]_1 \oplus |1]_2$ ~ "$q_1 \oplus q_2 = 0$" is distinguished from $|110] = |1]_1 \oplus |1]_2 \oplus |0]_3$ ~ "$q_1 \oplus q_2 \oplus \bar{q}_3 = 0$", where $\bar{q}_3$ explicitly indicates $q_3$ is missing from the left side: i.e. "$q_1 \oplus q_2 = 0$" means there are only two qubits involved, while "$q_1 \oplus q_2 \oplus \bar{q}_3 = 0$" means there are three qubits involved, but $q_3$ is missing in this particular condition.

Now with the 1-q-condition superposition defined in Equation (5) and multiple-q-condition interaction defined in Equation (6), we can easily define the entangled conditions as:

$$b_1 |u_1 v_1] + b_2 |u_2 v_2] \tag{8}$$

where $|u_1]$, $|u_2]$, $|v_1]$ and $|v_2]$ are 1-q-conditions with $|u_1]$ orthogonal to $|u_2]$, and $|v_1]$ orthogonal to $|v_2]$ (an inner product can be defined in exactly the same way as the usual quantum state space). In particular, a quantum condition equivalent to the Bell state exists:

$$\begin{aligned}\phi &= \frac{1}{\sqrt{2}}(|00] + |11]) = \frac{1}{\sqrt{2}}\left[\left(|0]_1 \oplus |0]_2\right) + \left(|1]_1 \oplus |1]_2\right)\right] \\ &= \frac{1}{\sqrt{2}}\left[\left(\frac{1}{\sqrt{2}}(|+]_1 + |-]_1) \oplus \frac{1}{\sqrt{2}}(|+]_2 + |-]_2)\right) + \left(\frac{1}{\sqrt{2}}(|+]_1 - |-]_1) \oplus \frac{1}{\sqrt{2}}(|+]_2 - |-]_2)\right)\right] \\ &= \frac{1}{\sqrt{2}}\left[\left(|+]_1 \oplus |+]_2\right) + \left(|-]_1 \oplus |-]_2\right)\right] = \frac{1}{\sqrt{2}}(|++] + |--])\end{aligned} \tag{9}$$

where $|+] = \frac{1}{\sqrt{2}}(|0]_1 + |1]_1)$ and $|-] = \frac{1}{\sqrt{2}}(|0]_1 - |1]_1)$. We see in Equation (9) that with $\oplus$ in place of $\otimes$, $\phi$ behaves exactly like a Bell state with quantum correlations in two different bases $\{|0], |1]\}$ and $\{|+], |-]\}$: this is a quantum-only behavior with no classical equivalent.

Having established the formal representation and physical interpretation of the quantum condition space, next we propose a simple example of a potential realization of the quantum conditions with quantum circuits. As previously explained in Section 2, given a quantum state vector represented by qubits, apply $\text{CNOT}_{1 \to 2}$ and the value of $q_1 \oplus q_2$ will be stored on $q_2$. Now to realize the product condition in Equation (6) we need three working qubits $q_1$, $q_2$ and $q_3$, and two ancilla qubits $q_{a1}$ and $q_{a2}$, as shown in Figure 1. Initialize $q_3 = |0\rangle$, $q_{a1} = a_1 |0\rangle_{a1} + a_2 |1\rangle_{a1}$, and $q_{a2} = b_1 |0\rangle_{a2} + b_2 |1\rangle_{a2}$. Now apply the Toffoli gates $\text{CCNOT}_{a_1, 1 \to 3}$ (where "$a_1, 1 \to 3$" means $q_{a1}$ and $q_1$ control $q_3$, same for $\text{CCNOT}_{a_2, 2 \to 3}$ next) and $\text{CCNOT}_{a_2, 2 \to 3}$, then the value of $(a_1 |0]_1 + a_2 |1]_1) \oplus (b_1 |0]_2 + b_2 |1]_2)$ will be stored on $q_3$, thus by measuring $q_3$ the quantum condition in Equation (6) has been realized on the quantum state space of $q_1$ and $q_2$. Similarly, to realize the entangled condition in Equation (8), just initialize $q_3 = |0\rangle$, $q_{a1} q_{a2} = b_1 |u_1 v_1\rangle + b_2 |u_2 v_2\rangle$, and then apply $\text{CCNOT}_{a_1, 1 \to 3}$ and $\text{CCNOT}_{a_2, 2 \to 3}$: now the value of $b_1 |u_1 v_1] + b_2 |u_2 v_2]$ will be stored on $q_3$ and the quantum condition in Equation (8) has been realized on the quantum state space of $q_1$ and $q_2$. These procedures can be used to create arbitrary quantum conditions involving $n$ q-conditions. This demonstrates the theory of the quantum condition space can be realized physically.

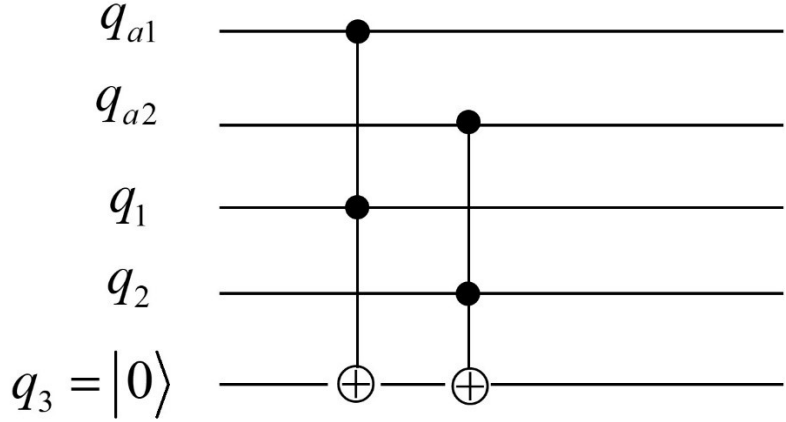

Figure 1. A physical realization of the quantum conditions. By initializing various quantum states on the ancilla qubits $q_{a1}$ and $q_{a2}$, and storing the condition value on $q_3$, any quantum condition can be realized on the space formed by $q_1$ and $q_2$.

### 3.3 Fourier transform and the entropic uncertainty principle

So far we have studied the following concepts:

a. The outcome space $V$ including vectors such as (001), (011) etc.
b. The quantum state space generated by using vectors in $V$ as basis states, including quantum states $\psi = \sum_{h=1}^{2^n-1} C_h |h\rangle$.
c. The dual space of the outcome space – the condition space $V^*$ including members such as "$q_1 \oplus q_2 = 0$", "$q_3 = 0$", etc.
d. The quantum condition space generated by using vectors in $V^*$ as basis states, including quantum conditions $\phi = \sum_{j=0}^{2^n-1} D_j |j]$.

Here the relation between $V$ and $V^*$ is analogous to the relation between the position variable $x$ and the momentum variable $k$ – the position and momentum are a pair of dual variables and quantum wavefunctions can be expressed in either the position representation $f(x)$ or the momentum representation $\hat{f}(k)$. Basic quantum physics says the two representations are related by a Fourier transform:

$$f(x) = \frac{1}{\sqrt{2\pi}} \int_{k-space} \hat{f}(k) e^{ikx} dk \qquad \hat{f}(k) = \frac{1}{\sqrt{2\pi}} \int_{x-space} f(x) e^{-ikx} dx \qquad (10)$$

The duality between position and momentum is an example of the mathematical concept of Pontryagin duality [24] that also applies to our $V$, $V^*$, and the quantum spaces generated by these two. $f(x)$ can be interpreted as the probability amplitude evaluated at $|x\rangle$ in the position space, and $\hat{f}(k)$ can be interpreted as the probability amplitude evaluated at $|k\rangle$ in the momentum space. The Fourier transform in Equation (10) thus relates the probability amplitude in the position space to that in the momentum space. Similarly, in our discrete case the $C_h$ in $\psi = \sum_{h=0}^{2^n-1} C_h |h\rangle$ can be understood as the probability evaluated at $|h\rangle$ in the outcome space, and the $D_j$ in $\phi = \sum_{j=0}^{2^n-1} D_j |j]$ can be understood as the probability evaluated at $|j]$ in the condition space. So a Fourier transform should exist to relate the $C_h$'s to the $D_j$'s. Define functions:

$$f(|h\rangle) = C_h, \quad \hat{f}(|j]) = D_j \tag{11}$$

then we have a Fourier transform:

$$f(|h\rangle) = \frac{1}{\sqrt{2^n}} \sum_{j=0}^{2^n-1} \hat{f}(|j]) E(h,j) \quad \hat{f}(|j]) = \frac{1}{\sqrt{2^n}} \sum_{h=0}^{2^n-1} f(|h\rangle) E(h,j) \tag{12}$$

where the function $E(h,j)$ is defined for $h$ in binary form $h = (h_1 h_2 ... h_n)$ and $j$ in binary form $j = (j_1 j_2 ... j_n)$ such that:

$$E(h,j) = \exp\left(i\pi \sum_{m=1}^{n} h_m j_m\right) \tag{13}$$

Using $n = 2$ as an example, from $f$ to $\hat{f}$ we have:

$$\begin{aligned}
\hat{f}(|00]) &= \frac{1}{2}\left[f(|00\rangle) + f(|01\rangle) + f(|10\rangle) + f(|11\rangle)\right] \\
\hat{f}(|01]) &= \frac{1}{2}\left[f(|00\rangle) - f(|01\rangle) + f(|10\rangle) - f(|11\rangle)\right] \\
\hat{f}(|10]) &= \frac{1}{2}\left[f(|00\rangle) + f(|01\rangle) - f(|10\rangle) - f(|11\rangle)\right] \\
\hat{f}(|11]) &= \frac{1}{2}\left[f(|00\rangle) - f(|01\rangle) - f(|10\rangle) + f(|11\rangle)\right]
\end{aligned} \tag{14}$$

which in matrix form is just the Walsh-Hadamard transform $H_2$ in $2^2$ dimensions. In fact, one can easily verify that this Fourier transform for any $n$ is always the Walsh-Hadamard transform $H_n$ in $2^n$ dimensions. We emphasize that this Fourier transform should not be confused with the quantum Fourier transform [26] commonly discussed in the quantum computing context.

In Equation (14) we notice that if a wavefunction is concentrated on a single $|j]$ then it is uniformly distributed over $|h\rangle$, and a simple inverse transform shows if it is concentrated on a single $|h\rangle$ then it is uniformly distributed over $|j]$: this result generalizes to all $n$, and an uncertainty principle should exist between the quantum state space and the quantum condition space. The usual form of the uncertainty principle between the position space and the momentum space is $\sigma_x \sigma_k \geq \frac{1}{2}$, where the $\sigma$ is the standard deviation. As our spaces are discrete we cannot use the standard deviation but instead should use the entropic uncertainty principle [27-29] such that:

$$H_S + H_C > 0$$
$$H_S = -\sum_{h=0}^{2^n-1} |f(|h\rangle)|^2 \log\left(|f(|h\rangle)|^2\right) \qquad H_C = -\sum_{j=0}^{2^n-1} |f(|j])|^2 \log\left(|f(|j])|^2\right) \qquad (15)$$

where $H_S$ and $H_C$ are the information entropies for the quantum state and quantum condition respectively. Equation (15) says the sum of the information entropies of the quantum state and the corresponding quantum condition is strictly greater than zero. The exact value of the lower bound should be a function of $n$ and will be derived in a future study. In addition to being useful for defining the uncertainty principle, the information entropies in Equation (15) may also have applications in quantum information. For example, in the context of quantum encryption using quantum states as the ciphertexts [30], we may be interested in maximizing the sum $H_S + H_C$ or balancing $H_S$ and $H_C$ such that the ciphertexts create maximal difficulty for an adversary Eve.

### 4. Relation between the half-set conditions and quantum circuits

The theory of the quantum condition space was originally motivated by the event probability evaluation in Section 2, where the Boolean algebra built by the half-set conditions is used to specify arbitrary events on the outcome space. In this section we discuss another potential application of the theory with a deeper relation between the half-set conditions and quantum circuits. In quantum computing any arbitrary quantum circuit is realized by a sequence of elementary quantum gates including 1-qubit unitaries $U_i$ acting on $q_i$ and 2-qubit controlled-unitaries $CU_{i \to j}$ using $q_i$ to control $q_j$. If we study the effects of the elementary gates on a quantum state vector space, for example a 4-dimensional space for two qubits:

$$U_1 = \begin{pmatrix} u_1 & 0 & u_2^* & 0 \\ 0 & u_1 & 0 & u_2^* \\ u_2 & 0 & -u_1^* & 0 \\ 0 & u_2 & 0 & -u_1^* \end{pmatrix} \quad U_2 = \begin{pmatrix} u_1 & u_2^* & 0 & 0 \\ u_2 & -u_1^* & 0 & 0 \\ 0 & 0 & u_1 & u_2^* \\ 0 & 0 & u_2 & -u_1^* \end{pmatrix}$$

$$CU_{1\to 2} = \begin{pmatrix} 1 & 0 & 0 & 0 \\ 0 & 1 & 0 & 0 \\ 0 & 0 & u_1 & u_2^* \\ 0 & 0 & u_2 & -u_1^* \end{pmatrix} \quad CU_{2\to 1} = \begin{pmatrix} u_1 & 0 & u_2^* & 0 \\ 0 & 1 & 0 & 0 \\ u_2 & 0 & -u_1^* & 0 \\ 0 & 0 & 0 & 1 \end{pmatrix}$$

(16)

where $|u_1|^2 + |u_2|^2 = 1$. We see in Equation (16) that the 1-qubit unitaries $U_1$ and $U_2$ each modifies the entries associated with the entire outcome space, while the controlled-unitaries $CU_{1\to 2}$ and $CU_{2\to 1}$ each only modifies the entries associated with a half-set of outcomes. This result is also true for a $2^n$ space for $n$ qubits: in general, any 1-qubit unitary modifies the entire quantum space spanned by all the outcomes, and any 2-qubit controlled-unitary modifies half of the quantum space spanned by a half-set of outcomes. Note that the $CU$ gates in Equation (16) by convention act on $|1\rangle$ of the control qubit, but if we consider those $CU$ gates that act on $|0\rangle$, then any 1-qubit unitary can be considered as a pair of $CU$ gates acting on $|0\rangle$ and $|1\rangle$ respectively. In other words, any elementary quantum gate can be considered as modifying the entries associated with a half-set of outcomes specified by a half-set condition. Therefore a natural relation between quantum gates and half-set conditions has been established. Now notice that in Equation (16) although $U_1$ modifies all the entries, it does not do so in a free way: the 1st and 3rd entries are modified by the 2-by-2 matrix $\begin{pmatrix} u_1 & u_2^* \\ u_2 & -u_1^* \end{pmatrix}$, while the 2nd and the 4th entries are modified by the same matrix $\begin{pmatrix} u_1 & u_2^* \\ u_2 & -u_1^* \end{pmatrix}$. If the space is enlarged by having more qubits, the same happens that the $2^n$ entries are divided into many subspaces of 2-dimensions modified by the same matrix $\begin{pmatrix} u_1 & u_2^* \\ u_2 & -u_1^* \end{pmatrix}$. This explains why 1-qubit gates alone cannot achieve universality, because they cannot introduce *free changes*. On the other hand, a 2-qubit $CU$ gate only acts on a half-set of entries, and thus can introduce free changes to this half-set as compared to its complement (note the changes within the half-set itself is still not free). Now consider the fact that universality can be achieved by including 2-qubit $CU$ gates in the elementary gate set. Then as universality requires that free changes on any arbitrary subset must be achievable with enough number of $CU$ gates, we conclude that a sequence of $CU$ gates effectively realizes the Boolean operations of "and", "or", and "not" on the half-set conditions, thus allowing us to modify any arbitrary subsets (events as in Section 2) with enough steps. Now consider the sequence of elementary gates used to realize a quantum circuit, it also defines a sequence of half-set conditions and certain Boolean operations on these conditions.

Therefore in terms of complexity, a polynomial quantum circuit (one realized by a sequence of elementary gates with the length scaled as a polynomial of qubit number $n$) must correspond to a subset (event) of outcomes that can be specified by a polynomial number of Boolean operations on half-set conditions. In other words, the half-set conditions corresponding to the $CU$ gates play an important role in quantifying the complexity of a quantum circuit. The structure of the half-set conditions and the Boolean operations define how much freedom is allowed in modifying the quantum space, and once the structure has been fixed, the actual parameters of the unitaries can only introduce limited freedom in modifying the quantum space. In this sense, any quantum circuit that has decisive quantum advantage over classical operations should modify the quantum state vector with limited number of free changes (specified by a polynomial structure of half-set conditions and Boolean operations) such that a quantum circuit realization is simple, but exponential number of unfree changes such that any classical realization is hard. As a promising future direction, a more strictly defined correspondence between the elementary gate sequence and the Boolean operations, and a more quantified measure of free/unfree changes, would allow us to better understand the complexity of quantum circuits and use this knowledge to design efficient quantum algorithms.

## 5. Conclusion

In this work we first proposed using projection measurements to efficiently evaluate event probability on a quantum state vector representing a probability distribution. We then proceeded to study what kinds of events can be specified by the half-set conditions such that they can be easily realized by quantum gates and measurements. We found that the half-set "0" conditions form a condition space that is the dual of the outcome space and thus a natural correspondence between the two spaces can be established. In the same way the outcome space generates the quantum state space, the condition space also generates the quantum condition space that is the central idea of this work. We then provided a formalism for the quantum condition space, a physical interpretation using the creation of the q-condition, and a potential realization with the usual tools of quantum circuits. Similar to the quantum state space, the quantum condition space permits the existence of entangled conditions that have no classical equivalent. Analogous to the way in which the position space is related to the momentum space, the quantum condition space is related to the quantum state space by a Fourier transform guaranteed by the Pontryagin duality, and therefore an entropic uncertainty principle can be defined on them. The quantum condition space offers a novel perspective of understanding quantum states with the duality picture. In addition, the quantum conditions have physical meanings and realizations of their own and thus may be studied for purposes beyond the original motivation of characterizing events for probability evaluation. Finally, a deeper relation between the half-set conditions and quantum circuits provides insights into how quantum states are collectively modified by quantum gates, which leads to future directions in better understanding of the complexity of quantum circuits.

**Acknowledgements**

ZH thanks Peng Zhou for mathematical discussions. ZH and SK acknowledge funding by the U.S. Department of Energy (Office of Basic Energy Sciences) under Award No. DE-SC0019215.

**References**

1. Georgescu, I.M., S. Ashhab, and F. Nori, *Quantum simulation.* Reviews of Modern Physics, 2014. **86**(1): p. 153-185.
2. Montanaro, A., *Quantum algorithms: an overview.* npj Quantum Information, 2016. **2**(1): p. 15023.
3. Cao, Y., et al., *Quantum Chemistry in the Age of Quantum Computing.* Chemical Reviews, 2019. **119**(19): p. 10856-10915.
4. Albash, T. and D.A. Lidar, *Adiabatic quantum computation.* Reviews of Modern Physics, 2018. **90**(1): p. 015002.
5. Preskill, J., *Quantum Computing in the NISQ era and beyond.* Quantum, 2018. **2**: p. 79.
6. Kais, S., ed. *Quantum Information and Computation for Chemistry*. Quantum Information and Computation for Chemistry. 2014, John Wiley & Sons.
7. Preskill, J., *Quantum computing 40 years later.* arXiv:2106.10522 [quant-ph], 2021.
8. Arute, F., et al., *Quantum supremacy using a programmable superconducting processor.* Nature, 2019. **574**(7779): p. 505-510.
9. Boixo, S., et al., *Evidence for quantum annealing with more than one hundred qubits.* Nature Physics, 2014. **10**(3): p. 218-224.
10. Linke, N.M., et al., *Experimental comparison of two quantum computing architectures.* Proceedings of the National Academy of Sciences, 2017. **114**(13): p. 3305.
11. Carolan, J., et al., *Universal linear optics.* Science, 2015. **349**(6249): p. 711.
12. Zhong, H.-S., et al., *Quantum computational advantage using photons.* Science, 2020. **370**(6523): p. 1460.
13. Gong, M., et al., *Quantum walks on a programmable two-dimensional 62-qubit superconducting processor.* Science, 2021. **372**(6545): p. 948.
14. Kitaev, A.Y., *Quantum computations: algorithms and error correction.* Russ. Math. Surv., 1997. **52**: p. 1191.
15. Shor, P.W., *Polynomial-Time Algorithms for Prime Factorization and Discrete Logarithms on a Quantum Computer.* SIAM J. Comput., 1997. **26**(5): p. 1484–1509.
16. Harrow, A.W., A. Hassidim, and S. Lloyd, *Quantum Algorithm for Linear Systems of Equations.* Physical Review Letters, 2009. **103**(15): p. 150502.
17. Peruzzo, A., et al., *A variational eigenvalue solver on a photonic quantum processor.* Nature Communications, 2014. **5**(1): p. 4213.
18. Daskin, A. and S. Kais, *Decomposition of unitary matrices for finding quantum circuits: Application to molecular Hamiltonians.* The Journal of Chemical Physics, 2011. **134**(14): p. 144112.

19. Biamonte, J., et al., *Quantum machine learning.* Nature, 2017. **549**(7671): p. 195-202.
20. Xia, R. and S. Kais, *Quantum machine learning for electronic structure calculations.* Nature Communications, 2018. **9**(1): p. 4195.
21. Hu, Z., R. Xia, and S. Kais, *A quantum algorithm for evolving open quantum dynamics on quantum computing devices.* Scientific Reports, 2020. **10**(1): p. 3301.
22. Wang, H., S. Ashhab, and F. Nori, *Quantum algorithm for simulating the dynamics of an open quantum system.* Physical Review A, 2011. **83**(6): p. 062317.
23. Hu, Z., et al., *A general quantum algorithm for open quantum dynamics demonstrated with the Fenna-Matthews-Olson complex dynamics.* arXiv:2101.05287, 2021.
24. Morris, S.A., *Pontryagin Duality and the Structure of Locally Compact Abelian Groups*. London Mathematical Society Lecture Note Series. 1977, Cambridge: Cambridge University Press.
25. Hu, Z. and S. Kais, *Characterization of Quantum States Based on Creation Complexity.* Advanced Quantum Technologies, 2020. **n/a**(n/a): p. 2000043.
26. Nielsen, M.A. and I.L. Chuang, *Quantum Computation and Quantum Information: 10th Anniversary Edition*. 2011: Cambridge University Press. 708.
27. Hirschman, I.I., *A Note on Entropy.* American Journal of Mathematics, 1957. **79**(1): p. 152-156.
28. Özaydin, M. and T. Przebinda, *An entropy-based uncertainty principle for a locally compact abelian group.* Journal of Functional Analysis, 2004. **215**(1): p. 241-252.
29. Coles, P.J., et al., *Entropic uncertainty relations and their applications.* Reviews of Modern Physics, 2017. **89**(1): p. 015002.
30. Hu, Z. and S. Kais, *A quantum encryption scheme featuring confusion, diffusion, and mode of operation.* arXiv:2010.03062 [quant-ph], 2020.